# A Model For the WMAP Anomalous Ecliptic Plane Signal

H.N. Sharpe

Bognor, Ontario, Canada
sh3149@brucetelecom.com

## ABSTRACT

A simple model is presented to explain the high Galactic latitude anomalies in the WMAP data recently reported by Diego et al (2009). It is suggested that the anomalous deviation from a thermal spectrum could be caused by the propagation of background thermal radiation through a foreground optically thin HII cloud. The background radiation may be the remnant of cooling radio lobes associated with once-active jets from Sgr A*.

## INTRODUCTION

This note presents a model to explain the anomalous WMAP signal recently reported by Diego et al (2009). The model proposes that the observed anomaly consists of the combination of thermal radiation from a background discrete source plus a foreground thermal, free-free Bremssstrahlung radiation source from a typical optically thin HII region. The background source is tentatively identified with weakly radiating remnant cooling radio lobes associated with sporadic activity from the Sgr A* jets at the Galactic center. We make this association because the axis of the SW-NE anomalies reported by Diego has a similar inclination to the Sgr A* jet axis ( Falcke et al 2009).

**MODEL** (all quantities cgs)

Following Rybicki and Lightman (1979), the general equation for the observed brightness spectrum, $I_v(\tau)$ from a background source, $I_v(0)$ which passes through a foreground emitting/absorbing region with source function $S_v$ and optical depth $\tau_v$ is:

$$I_v^{obs}(\tau_v) = S_v + e^{-\tau_v}(I_v(0) - S_v) \qquad (1)$$

We will show later that the HII cloud optical depth $\tau_v \ll 1$ for WMAP frequencies. In this case, (1) becomes:

$$I_v^{obs}(\tau_v) = S_v + (1 - \tau_v)(I_v(0) - S_v)$$

$$\qquad\qquad = I_v(0) + \tau_v(S_v - I_v(0)) \qquad (2)$$

The optical depth, $\tau_\nu$ for thermal free-free absorption is:

$$\tau_\nu = 0.018 * n_e^2 * g_{ff} * L * T_{cloud}^{-3/2} * \nu^{-2} \tag{3}$$

where $n_e$ is the electron density, assumed equal to the ion density for the ionized hydrogen plasma, $g_{ff}$ is the gaunt factor and L is the line-of-sight cloud thickness.

For typical HII regions, $n_e = 10 \text{cm}^{-3}$, $g_{ff} = 1$, $T_{cloud} = 10^4 K$.

For these values we have:

$$\tau_\nu = 1.8 * 10^{-6} * L / \nu^2 \tag{4}$$

At WMAP frequencies $\tau_\nu \ll 1$ for realistic values of L confirming our earlier assumption that the HII cloud is optically thin.

We next use the Rayleigh-Jeans law for $I_\nu(0)$ and $S_\nu$:

$$I_\nu(0) = (2k/c^2) * T_{lobe} * \nu^2 \quad \text{and} \quad S_\nu = (2k/c^2) * T_{cloud} * \nu^2 \tag{5}$$

where $T_{lobe} = T_{cmb} + \Delta T_{lobe}$ is the effective temperature for the radiating jet lobe from Sgr A* and $T_{cloud}$ is the HII cloud temperature.

Substituting (5) and (4) in (2) we obtain:

$$I_\nu^{obs} = (2k/c^2) * [T_{lobe} * \nu^2 + 1.8*10^{-6}*L*(T_{cloud} - T_{lobe})] \tag{6}$$

For $T_{cloud} = 10^4 \gg T_{lobe}$ (see later) and expressing L in parsecs we have:

$$I_\nu^{obs} = (2k/c^2) * [T_{lobe} * \nu^2 + 5.4*10^{16}*L] \tag{7}$$

Diego et al (2009) presented a model for the anomaly which required an overbrightness $\Delta I_\nu = 10$ KJy/sr ($10^{-19}$ ergs cm$^{-2}$ sec$^{-1}$ Hz$^{-1}$ sr$^{-1}$) at 61 GHz. We use this value in (7) as a constraint for computing the required value for $\Delta T_{lobe}$:

$$\Delta I_\nu = (2k/c^2) * \Delta T_{lobe} * \nu^2$$

For this constraint we compute from (7) that $\Delta T_{lobe} = 100 \ \mu K$.

This is the required value for the lobe temperature above the CMB background temperature and suggests that the lobe is effectively at equilibrium with the CMB.

Next we compute the required value for the HII cloud line-of-sight thickness, L (parsecs) to satisfy the 10 $\mu$ K temperature anomaly reported by Diego for the WMAP linear combination V+W-2Q where: $v_V$ =61 GHz, $v_W$ =94 GHz and $v_Q$ =41 GHz.

In (7) we use:

$$I_v^{obs} = (2k/c^2) * v^2 * T_v^{obs} \tag{8}$$

and then form an expression for the observed temperature anomaly by Diego.

From (7) and (8):

$$T_v^{obs} = T_{lobe} + 5.4*10^{16}*L*v^{-2} \tag{9}$$

Then we have:

$$\Delta T_{V+W-2Q}^{obs} = (5.4*10^{16}*L)*[\, v_V^{-2} + v_W^{-2} - 2v_Q^{-2} \,] \tag{10}$$

Setting $\Delta T = 10 \mu$ K, we compute:

$$L = 0.23 \text{ pc}$$

This is the required HII cloud depth to match the 10 $\mu$ K temperature anomaly.

**SUMMARY**

We have shown that by postulating a distant weak thermal source of radiation with an effective observed temperature of 100 $\mu$ K above the CMB, and which passes through a typical HII optically thin region of line-of-sight depth 0.23 pc, it is possible to explain the WMAP temperature anomaly reported by Diego and the associated flux overdensity at 61 GHz.

Equation (7) shows that the thermal spectral power law 2 is not fundamentally changed in this proposed process. Rather, a constant term ( $5.4*10^{16}*L$ ) is added to the thermal spectrum from the propagation of the background lobe radiation through the HII radiation. This constant term slightly modifies the observed brightness spectrum at the WMAP frequencies from a true thermal spectrum. This modification originates in the inverse frequency dependence of the HII cloud optical depth.

Finally for the HII cloud parameters at WMAP frequencies it can readily be shown that the dominant opacity source is free-free Bremsstrahlung relative to Thompson scattering and inverse Compton scattering.

# REFERENCES

Diego, J.M et al. WMAP Anomalous Signal in the Ecliptic Plane. *arXiv*:0901.4344v1 [astro-ph.CO] 27 Jan 2009.

Falcke, H. et al. Jet-lag in SgR A*: What size and timing measurements tell us about the central black hole in the Milky Way. *arXiv*:0901.3723v1 [astro-ph.GA] 23 Jan 2009.

Rybicki, G.B. and A.P. Lightman. *Radiative Processes in Astrophysics*. Wiley 1979.